\documentclass{iauc}
\usepackage{graphicx}

\title{The Faint End of the HI Mass Function}
\author{K. Kova\u{c}$^1$,
  T.A.Oosterloo$^2$\break
\and  J.M.van der Hulst$^1$}

\affiliation{$^1$Kapteyn Astronomical Institute, P.O. Box 800, 9700 AV
	   Groningen, The Netherlands \break
	   email:kovac@astro.rug.nl\\[\affilskip] $^2$ASTRON,
	   Dwingeloo, The Netherlands}
\pubyear{2005}
\volume{198}  %% insert here IAU Symposium No.
\pagerange{1-4}
\date{?? and in revised form ??}
\jname{Near--Field Cosmology with Dwarf Elliptical Galaxies}
\editors{B.Binggeli, H.Jerjen, eds.}
\begin{document}
\def\HI{H\,{\small I}}
\maketitle
\begin{abstract}

We study the faint end of the HI mass function (HIMF) in order to test
the predictions of the CDM theory on the number density of objects
with small (dark) masses. The neutral hydrogen is much better tracer
of the underlying mass distribution compared to the luminous matter
and can be used to test the existence of a population of small
galaxies in which the star formation has been partially or completely
suppressed during cosmic evolution. Due to technical limitations, the
existing HI surveys are not very sensitive to HI masses below 10$^8$
$M_{\odot}$.  We designed a blind HI survey to be sensitive to
objects with small HI masses. The surveyed area is in the Canis
Venatici groups of galaxies and covers in total $\approx$ 86 deg$^2$
of sky, with observed velocities in the range $-350 < cz < 1400$
km s$^{-1}$. We detected 69 objects, 22 of them for the first time in
HI. All new HI detections fall in the lower part of the
mass-histogram, confirming our ability to detect galaxies with small
HI masses. The calculated HIMF is flat in the faint end regime
($\alpha \sim$ -1), different from the steep rise predicted by CDM
models. Possible effects of the environment on the estimated HIMF
parameters are discussed.

\end{abstract}

\section{Motivation}
Some works that address the discrepancy between the number of the
observed dwarf galaxies in the optical and the number of those
galaxies predicted by CDM models, suggest that small galaxies contain
relatively less luminous matter than larger galaxies
(e.g. \cite[Verde, Oh \& Jimenez 2002]{ver}). If that is the case, a
fraction of galaxies with dark matter haloes of mass below 10$^{9}$ -
10$^{10}$ $M_{\odot}$ may never form stars and instead they will remain
dark. Neutral hydrogen, if still within these systems, would be a much
better tracer of dark galaxies. If one assumes that HI makes up a
few percent of the total mass of a galaxy, dark galaxies would contain
HI in the range 10$^7$ - 10$^8$ $M_{\odot}$ or less. The surveys
carried out in HI have poor statistics for objects with masses below
10$^8$ $M_{\odot}$. Moreover, all HI objects detected on extragalactic
scales have an optical counterpart and have masses above 10$^7$
$M_{\odot}$ \cite[(Briggs 2004)]{bri}. On the other hand, there is a
population of recently discovered compact HI clouds (CHVCs) around the
Milky Way and M31 \cite[(de Heij et al. 2002)]{deH}. The CHVCs have HI
masses below 10$^6$ $M_{\odot}$ and none of them has an optical
counterpart. The gap in the observed HI masses between 10$^6$ - 10$^7$
$M_{\odot}$ is due to technical limitation of existing surveys. The 
existing HI information is not sufficient in the
range of HI masses predicted in these galaxies with little
or no stars. There is obviously a need for an HI survey in which the 
galaxies with HI masses below 10$^8$ $M_{\odot}$ are a significant 
fraction of all detections.

\section{The CVn blind survey}

We designed an HI survey which is extremely sensitive for objects with
HI masses below 10$^8$ $M_{\odot}$. The main difficulty for the HI
observations are the long integration times. To be able to detect a
statistically meaningful sample of objects with low HI masses in a
reasonable amount of time, we have chosen to observe in a slightly
over-dense region instead of surveying the field. The Canis
Venatici (CVn) groups of galaxies are an excellent region for this
purpose. The CVn galaxies extend along the line of sight up to a
distance determined by $cz \approx 1200$ km s$^{-1}$. This region
contains a large number of the dwarf galaxies, rich in HI, as detected
in the previous surveys (in the optical by \cite [Binggeli et
al. 1990]{bin}, in the HI by \cite[Kraan-Korteweg et al. 1999]{kra}).

For the survey, we observed 60 $\times$ 12 hr using the Westerbork
Synthesis Radio Telescope (WSRT) in mosaic mode. The observed area on
the sky is $\approx$ 86 deg$^2$. It is covered by observing 60 strips
of different declination, separated by 15 arcmin. Each strip of
constant declination has 24 pointings separated by 15 arcmin in Right
Ascension. The total integration time per pointing was 80.1 min. Only data
with $-350 < cz < 1400$ km s$^{-1}$ were used in further data
reduction and analysis process. Given that the size of the WSRT
primary beam is 34 arcmin, we obtained
nearly uniform sensitivity over the whole observed area. This makes the
CVn survey extremely suitable to detect objects with small HI masses.

The data were calibrated and analysed using scripts written by
Oosterloo and Kova\u{c} based on the the MIRIAD (\cite[Sault at
al. 1995]{sau}) routines. The output of this process is $\sim$ 1400
3-dimensional data-cubes, each cube being created by extracting the
central part of the combined 9 nearest pointings. The spatial
resolution of the data-cubes produced is $\sim$ 30 $\times$ 60 arcsec.
The velocity resolution is $\sim$ 33 km s$^{-1} $. The noise in the
cubes is just below 0.9 mJy Beam$^{-1}$. The data-cubes were
automatically searched for signal, using algorithms implemented by
Kova\u{c}. The performance of the algorithms has been tested by
inserting simulated objects into the real data-cubes.  The HI mass of
the detections was estimated using $ M_{HI} = 236 D^2 \int S dv
[M_{\odot}] $ , where $D$ is the distance to the object in Mpc, $S$ is
the flux in mJy integrated over the velocity range with significant
detection, and the width of the channels is $dv$ given in km s$^{-1}$.
%At this moment the largest uncertainty in the estimated HI masses of
%the detected objects is their distance. 

\begin{figure}
\centering
\resizebox{6.0cm}{!}{\includegraphics{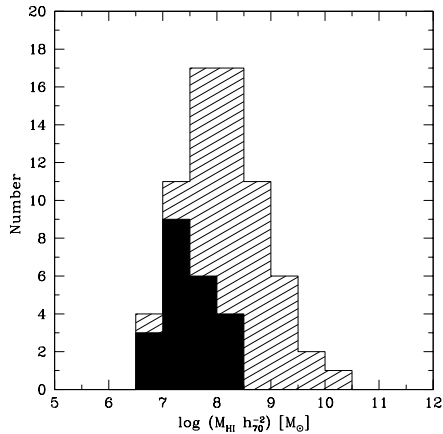}}
\caption{Number of detections per HI mass bin. The shaded histogram
represents the distribution of the whole sample of 69 detections, the
filled histogram represents the distribution of 22 detections observed
for the first time in HI.}

\end{figure}

We used the minimum detectable integrated flux to estimate the maximum
distance up to which the object with a certain HI mass can be detected
in our survey. Using the real and the simulated detections we
calculated that $S_{\rm min} = 0.2\ {\rm Jy\ km\ s}^{-1}$ is the
minimum detectable integrated flux to classify a detection in our
survey as a real object. The minimum HI mass which an object must have to be
detected in our survey is then given by $\log M_{{\rm HI,min}} = 4.67
+ 2 \log D [M_\odot]$ . Objects with HI masses 10$^6$ $h_{70}^{-2}$
$M_{\odot}$ can be detected up to a distance 4.6 $h_{70}^{-1}$ Mpc and
with HI masses 10$^7$ $h_{70}^{-2}$ $M_{\odot}$ up to 14.6 $h_{70}^{-1}$
Mpc.

In total, we detect 69 objects with HI emission. The cross-correlation
with objects from the existing catalogues has been done using the
astronomical data bases NED 
(http://nedwww.ipac.caltech.edu) and HyperLeda (http://leda.univ-lyon1.fr). 
We found that 4 of our HI
detections have not been previously catalogued and 1 object
does not have an optical counterpart. The last detection is most
probably a tidal satellite. It is important to point out that 22 of
our 69 detections have not been previously detected in HI. The number
distribution of detections per HI mass bin is shown in Figure 1. 

At this moment the largest uncertainty in the estimated HI masses of
the detected objects is their distance. We obtained the distances by
measuring recessional velocities and correcting them for the motion 
of the Sun in the Galaxy, motion of the Galaxy in the Local group 
and for the infall of the Local Group towards Virgo, using the infall 
velocity of the Local Group according to \cite[Theureau et al. (1998)]{the}.

\section{The HI Mass Function}

The HI mass function (HIMF) is defined analogously to the optical
luminosity function, so one can use identical methods for its
derivation. We used the $\Sigma(1/V_{max})$ method introduced by
\cite[Schmidt (1986)]{sch}. This method consists of summing reciprocal
values of the maximum surveyed volume in which the objects of a
certain HI mass can be detected.

\begin{figure}[h]
\centering
\resizebox{6.0cm}{!}{\includegraphics{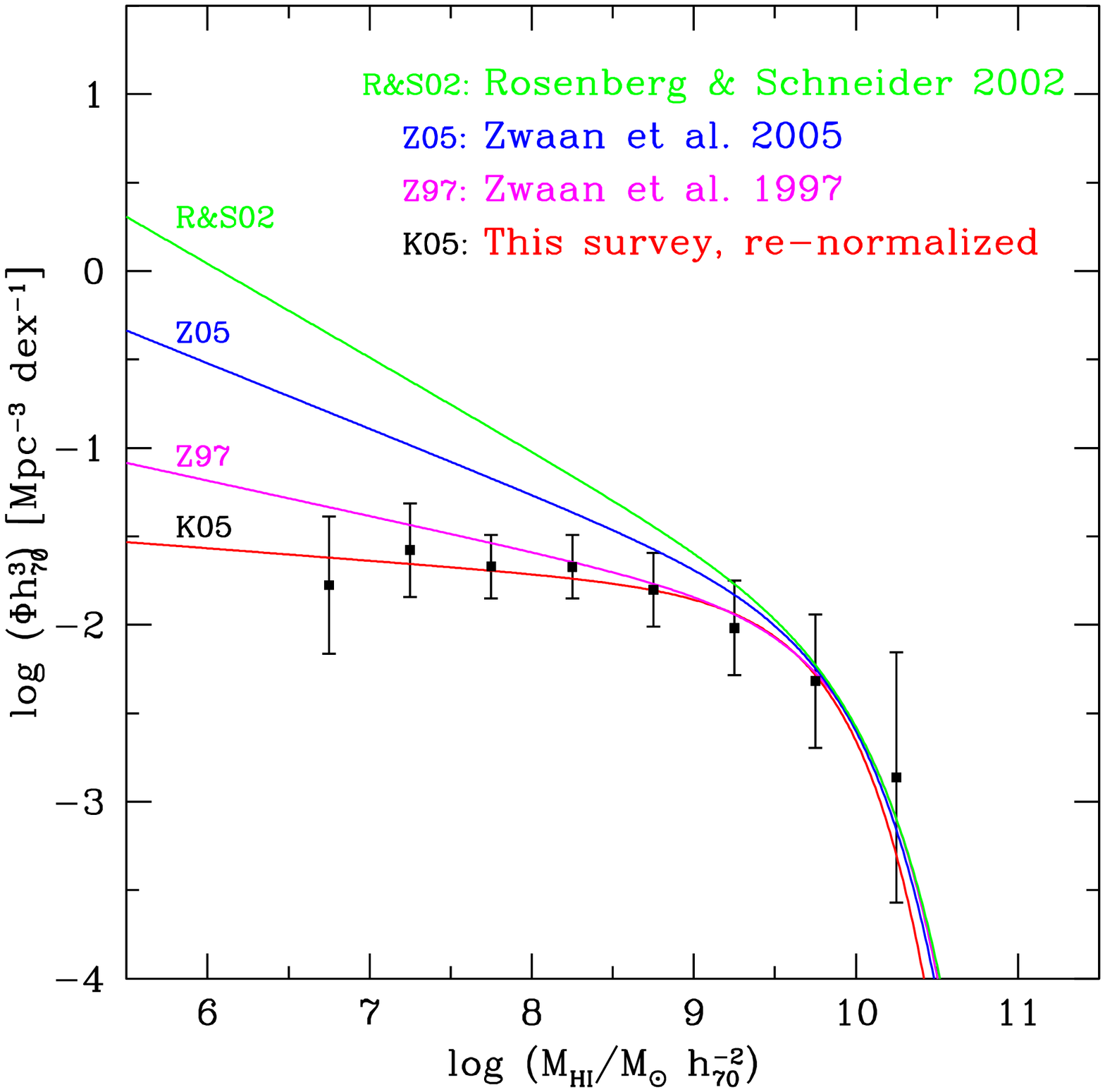}}
\resizebox{6.0cm}{!}{\includegraphics{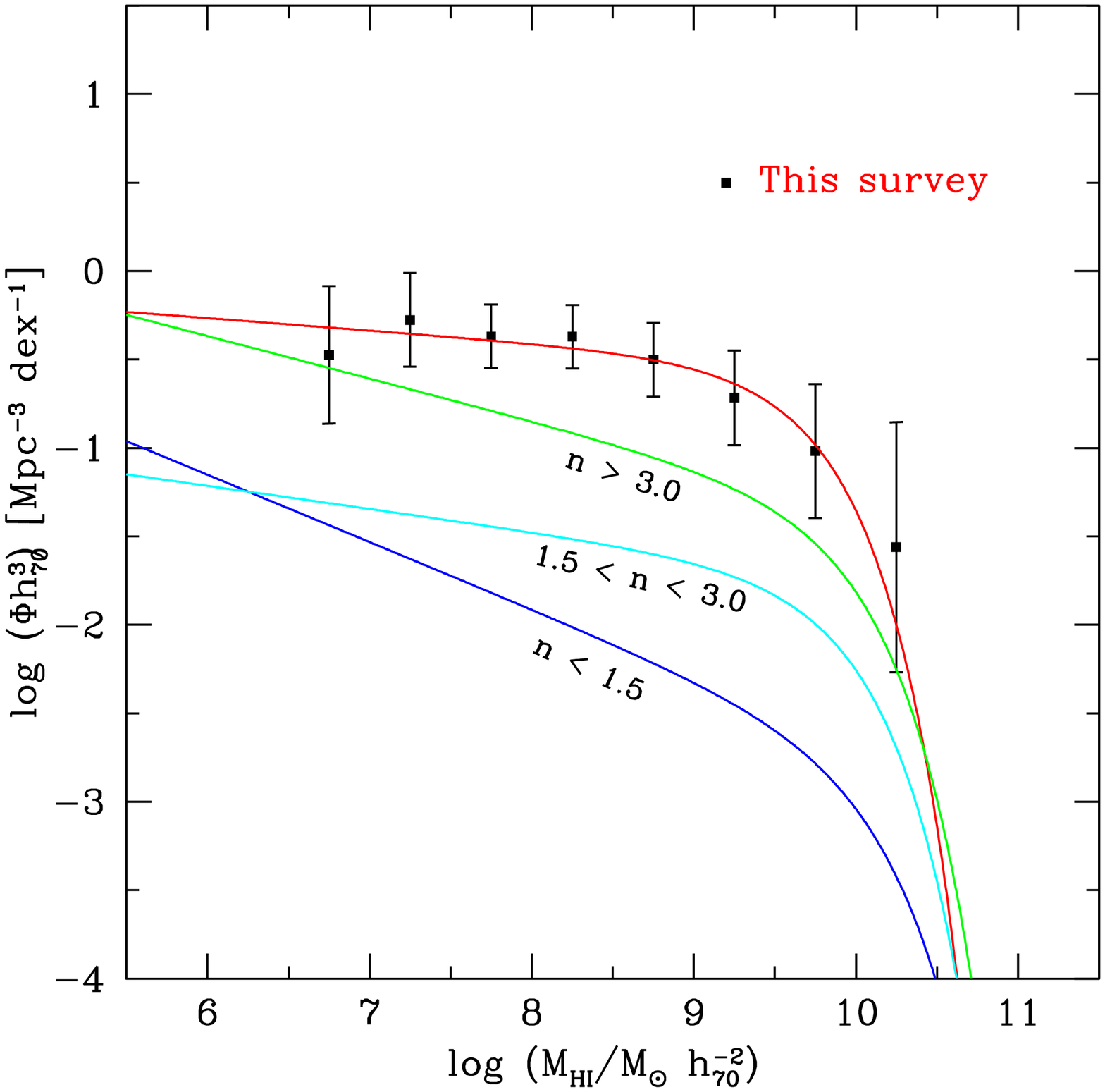}}

\caption{{\bf Left panel:} The HIMF from our survey compared to the
HIMFs derived from three recent blind HI surveys. Our function is
re-normalized to match the high mass end of the HIMF determined from
the plotted surveys. {\bf Right panel:} Comparison of the HIMF from our
survey to the HIMF derived for regions of three different local matter
densities. The higher values of $n$ correspond to regions of higher
density (\cite[Springob et al. 2005]{S05}). }
\end{figure}

We fitted the binned data with a Schechter function. The best-fit
values are $\alpha = -1.07$, $\log(M^* / M_{\odot}) = 9.73$ and $\phi
= 0.129 h_{70}^{3}$ Mpc$^{-3}$. The calculated HIMF is presented in
Figure 2. In the left panel of Figure 2, the HIMF from this survey
is re-normalized to match the the high mass end of the HIMF derived
from three recent blind HI surveys. The real number density of the
HI objects is 20 $\times$ higher than plotted. For
the surveyed volume, the number density of the objects with HI masses
spanning $\sim 2$ decades (10$^7$ - 10$^9$ in $M_{\odot}$) is
approximately constant. This result is not consistent with the
slopes of the HIMF as found by \cite[Zwaan et al (2005)]{Z05} and
\cite[Rosenberg \& Schneider (2002)]{RS02}. The disagreement is even
larger when comparing this slope to the analytically derived slope
$\sim$ -1.8 of the low mass end of the CDM mass function \cite[(Press
\& Schechter 1974)]{ps}.

Recently, a lot of effort has been spent to explain the effect of
environment on the properties of galaxies and their distributions. The
HI surveys cover a much smaller area than the optical surveys. The
only way to study the environmental effects on the HIMF is by direct 
comparison of
the estimated HIMF from the surveys carried out in different
environments (clusters, groups, voids). The main conclusion from this
comparison is that HI galaxies detected in a denser environment have
HIMF characterised with the flatter end compared to the HIMF derived
from the field. \cite[Springob et al. (2005)]{S05} investigated the HIMF
of optically selected HI galaxies. Their sample was large enough to investigate
the dependence of the HIMF on the environment. Since their
subsamples of galaxies in regions of different densities had
 the same fractions of morphological types, the
difference in the derived HIMF must be due to the density of the
environment. Although the results obtained had statistically small
differences, the conclusion was that the HIMF in the higher-density
region shows a flattening in the faint end and lower values of
$M_*$. Thus, the result of the HIMF from our survey is in agreement
with this finding  (Figure~2, right panel).

\begin{acknowledgements} The Westerbork Synthesis Radio Telescope is
operated by the ASTRON (Netherlands Foundation for Research in
Astronomy) with support from the Netherlands Foundation for Scientific
Research (NWO). K.K. is grateful to the IAU for the travel
grant to attend this conference. 

\end{acknowledgements}

\begin{discussion}

\discuss{Karachentsev} {About 50 $\%$ of the area
surveyed by you lies within the Arecibo zone. This area will be
surveyed soon with much deeper detection limit. It would be reasonable
to shift your strip of survey at higher declination, i.e. outside the
Arecibo zone.}

\end{discussion}


\begin{thebibliography}{}
\bibitem[Binggeli \etal 1990]{bin}
{Binggeli, B., Tarenghi, M. \& Sandage A.} 1990, \textit{A\&A} {228}, 42 
\bibitem[Briggs (2004)] {bri}
{Briggs, F.H.} 2004, \textit{IAUS} {217}, 26
\bibitem[de Heij \etal (2002)]{deH} 
{de Heij, V., Braun, R. \& Burton, W.B.} 2002, \textit{A\&A} {392}, 417
\bibitem[Kraan-Korteweg \etal 1999]{kra} {Kraan-Korteweg, R.C., van
Driel, W., Briggs, F., Binggeli, B., Mostefaoui, T. I.} 1999,
\textit{A\&AS} {135}, 255
\bibitem[Press \& Schechter 1974]{ps}
{Press, W.H. \& Schechter, P.} 1974, \textit{ApJ} {187}, 425
\bibitem[Rosenberg \& Schneider (2002)]{RS02}	
{Rosenberg, J.L. \& Schneider, S.E.} 2002, \textit{ApJ} {567}, 247
\bibitem[Sault \etal 1995]{sau} {Sault, R.J., Teuben, P.J. \& Wright,
M.C.H.} 1995, in:\textit{ASP Conf.Ser. 77: Astronomical Data Analysis
Software and Systems IV.} ,433
\bibitem[Schmidt (1986)]{sch}
{Schmidt, M.} 1986, \textit{ApJ} {151}, 393
\bibitem[Springob \etal 2005]{S05}
{Springob, C.M., Haynes, M.P. \& Giovanelli, R.} 2005, \textit{ApJ} {621},215
\bibitem[Theureau \etal (1998)]{the} {Theureau, G., Rauzy, S.,
Bottinelli, L. \& Gouguenheim, L.} 1998, \textit{A\&A} {340}, 21
\bibitem[Verde, Oh, \& Jimenez 2002] {ver}
 {Verde, L., Oh, S.P. \& Jimenez, R.} 2002, \textit{MNRAS} 336, 541
\bibitem[Zwaan \etal 1997]{Z97}
{Zwaan, M.A., Briggs, F. H., Sprayberry, D. \& Sorar, E.} 1997,
\textit{ApJ} {490}, 173 
\bibitem[Zwaan \etal 2005]{Z05} {Zwaan, M.A.,
Meyer, M.J., Staveley-Smith, L. \& Webster, R.L.} 2005, \textit{MNRAS} {359}
L30
 \end{thebibliography}
\end{document}